\documentstyle[prl,aps,epsf]{revtex}
\begin{document}
\draft

\flushbottom
\twocolumn[
\hsize\textwidth\columnwidth\hsize\csname @twocolumnfalse\endcsname

\title{Finite-size scaling 
and conformal anomaly of the Ising model in curved space}
\author{J. Gonz\'alez \\}
\address{
        Instituto de Estructura de la Materia.  Consejo Superior
de Investigaciones Cient{\'\i}ficas.  Serrano 123, 28006 Madrid.
Spain.}
\date{\today}
\maketitle
\begin{abstract}
We study the finite-size scaling of the free energy of the Ising
model on lattices with the topology of the tetrahedron and the
octahedron. Our construction allows to perform changes in the
length scale of the model without altering the distribution
of the curvature in the space. We show that the subleading 
contribution to the free energy follows a logarithmic dependence, 
in agreement with the conformal field theory prediction. The 
conformal anomaly is given by the sum of the contributions
computed at each of the conical singularities of the space,
except when perfect order of the spins is precluded by 
frustration in the model.

\end{abstract}

]
\narrowtext 
\tightenlines
The introduction of conformal field theory about fifteen years
ago can be considered one of the most important developments
towards the understanding of critical phenomena in two
dimensions\cite{bpz}.  
This subject added to the progress achieved ten
years before by application of the renormalization group ideas
to critical phenomena. Both the renormalization group and
conformal field theory have in common the idea of 
scaling\cite{cardy}. This
plays a central role in the confrontation with experimental
measurements near a critical point, as well as in numerical
simulations where the critical behavior is approached by
enlarging the size of the system\cite{barber}.

The influence of the gravitational background on
critical phenomena is largely unknown, though. This problem can be
approached from the point of view of conformal field theory,
which is able to deal with two-dimensional backgrounds related
by conformal transformations to the plane. Although the kind of
geometries one can handle in this way is restricted, we have
learned about the interesting properties of conformal field
theories on semiplanes\cite{semi}, cylinders\cite{cyl} and 
conical singularities\cite{peschel}.

In the case of a two-dimensional smooth manifold, 
it has been shown on general grounds
that the free energy $F$ has corrections that depend
directly on the central charge $c$ of the conformal field
theory\cite{cardy}. 
As a function of the length scale $L$ of the system, the
free energy has to behave in the form
\begin{equation}
F(L) = a L^2  + b \log (L) +  \ldots
\label{fe}
\end{equation}
where $b = - c \chi /12$ for a manifold with Euler 
characteristic $\chi $. 
A similar formula applies to the case of
a conical singularity\cite{peschel}. 
Logarithmic corrections to the free energy also arise 
associated to corners in higher-dimensional 
spaces\cite{priv1} 
Until now,
though, no examples of statistical models have been
considered where the 
logarithmic corrections due to the curvature have been tested.
The question is actually nontrivial since, as we will see
below, the simplest lattices that make feasible the
construction of models with such scaling behavior do not
give rise to smooth manifolds in the continuum limit.

\begin{figure}
\begin{center}
\epsfysize=4cm
\mbox{\epsfbox{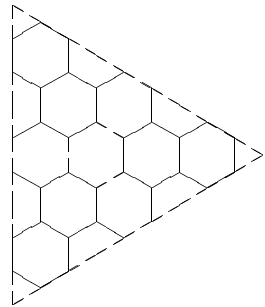}}
\end{center}
\caption{A generic triangular block for the lattices embedded on
the tetrahedron and the octahedron.}
\label{facet}
\end{figure}

In this paper we study the Ising model on lattices 
whose continuum limits have
the topology of the tetrahedron and the octahedron. Our aim is
to discern whether an expression like (\ref{fe}) applies,
providing then a check of the conformal field theory description
on the curved background. To accomplish this task we take the
thermodynamic limit along a series of honeycomb lattices which
are built by assembling triangular pieces like that in Fig.
\ref{facet} as the facets of the given polyhedron. Our choice of
this kind of lattices is determined by the feasibility of
growing them up regularly while preserving the geometry of the
polyhedron.  From the point of view of the simplicial geometry,
the local curvature at each $n$-fold ring of the lattice is
given by $R_i = \pi (6 - n_i)/n_i$. Thus, our honeycomb lattices
embedded on the tetrahedron, as well as in the octahedron, keep
the same distribution of the curvature (nonvanishing only at the
three-fold and four-fold rings that encircle the vertices of the
respective polyhedra), no matter the size of the lattice.

The critical behavior of the Ising model on the tetrahedron has
been discussed in Ref. \onlinecite{nos}. It has been shown
there that the critical exponents $\alpha , \beta $ and
$\gamma $ do not deviate in the curved geometry from the known
values of the Ising model on a flat space. In the present
Letter we focus on the imprints that the curvature may leave
in the scaling behavior of the statistical system. 
The $\log (L)$  correction to
the free energy is known in the case of a conical
singularity on a two-dimensional surface\cite{peschel}. 
By measuring
the $\log (L)$ scaling of the free energy we are then making a
nontrivial check of the conformal field theory prediction, since
we are dealing with geometries that are not small perturbations
with respect to flat space. This may also validate our construction
as an alternative procedure to the determination of the 
conformal anomaly in spaces with the topology of the 
cylinder\cite{wzw,vertex,alc,fss1,on,triang,xy}.

We begin by analyzing the Ising model
on the octahedron. Contrary to the case of
the tetrahedron, where there is frustration of the
model, the present lattices are bipartite since they are
built by assembling four of the
pieces in Fig. \ref{facet} around each vertex. 
The partition function has
to be then symmetric under the change of sign of the coupling
constant $\beta \rightarrow - \beta $. 

The thermodynamic limit has to be taken in order
to approach the critical point of the model. This may pose a
problem as long as we want to measure a scaling behavior like
(\ref{fe}) that is supposed to be well-defined at the critical
point of the continuum model. On a lattice of finite length
scale $L$, we may only define a ^^ ^^ pseudocritical''
coupling $\beta_L $ at which some of the observables, like for
instance the specific heat, reaches a maximum value\cite{barber}. 
We will discern
later what is the correct choice to extract
the genuine scaling of the conformal field theory
from the finite-size data.

From the technical point of view,
high-precision measurements are needed in order to observe
neatly a $\log (L)$ dependence of the free energy, after
subtraction of the leading contribution $\propto L^2 $. The dimer
approach affords such possibility, by translating the
computation of the partition function into the evaluation of the
Pfaffian of an antisymmetric operator $A$\cite{montr,mccoy}. 
This is given by a
coordination matrix of what is called the decorated lattice,
which is obtained in our case
by inserting a triangle in place of each of
the points of the original lattice. A detailed example of how to
build the coordination matrix for planar lattices similar to
ours can be found in Ref. \onlinecite{nos}. The partition
function can be written in the form
\begin{equation}
Z = (\cos \beta )^l \left( det(A) \right)^{1/2}
\label{z}
\end{equation}
where $l$ is the number of links of the original lattice.  From
(\ref{z}) it is clear that the partition function and the free
energy $F = - \log (Z) $ can be computed with high precision on
reasonably large lattices, as far as the evaluation of the
corresponding determinant becomes feasible. 

In the case of our lattices in curved space, the determination
of the logarithmic correction to $F$ is facilitated by the fact
that finite-size scaling sets in at very small lattice size.
The honeycomb lattices embedded on the octahedron form a family
with increasing number of sites $n = 24 N^2 , N = 1, 2,
\ldots $  The pseudocritical couplings 
approach the critical coupling $\beta_{\infty } =
\log (2 + \sqrt{3} )/2 \approx 0.6585 $ following the 
finite-size scaling law
\begin{equation}
\left| \beta_N - \beta_{\infty } \right| \sim 1/N^{\lambda }
\label{l}
\end{equation}
Usually $\lambda $ is related to the critical exponent $\nu $ of
the correlation length, $\lambda = 1/\nu $. One can check,
however, that in the case of the octahedron $\lambda $ is
sensibly higher than the expected value $\nu^{-1} = 1$.
The values of $\beta_N $, which we have computed by looking at the
maxima of the specific heat for $N = 2$ up to $N = 7$ 
(1176 lattice sites), are given in Table \ref{t1}.
By carrying out a sequence of fits, taking four consecutive
lattices for each of them, we obtain the respective estimates 
of the exponent $\lambda_{\rm octa} $, in order of increasing 
lattice size :  1.825, 1.809, 1.798, 1.794.

\begin{figure}
\begin{center}
\epsfysize=7cm
\mbox{\epsfbox{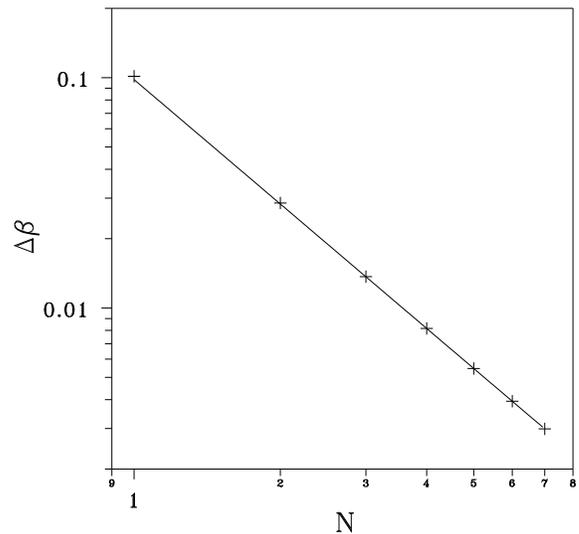}}
\end{center}
\caption{Deviation of the pseudocritical couplings from $\beta_{\infty }$
versus the length scale $N$.}
\label{lambda}
\end{figure}

We present in Fig. \ref{lambda} a logarithmic plot of the values
of $ \beta_N - \beta_{\infty } $ versus $N$ and the linear
fit for the last four points. It is remarkable the small
deviation of the points from the law (\ref{l}), even for the
smaller lattices, which ensures that the estimates for
$\lambda_{\rm octa} $ are converging to a value different to that
expected in flat space. The exponent for
the octahedron is very close to the exponent obtained in the
case of the tetrahedron, for a wider range of lattice sizes,
$\lambda_{\rm tetra} \approx 1.745(2) $ \cite{nos}. 
We recall that these estimates do not point at
a critical exponent of the correlation length different from
$\nu = 1$, but rather at a violation of the Ferdinand-Fisher
criterion for the determination of $\nu $ in the curved spaces.

\begin{figure}
\begin{center}
\epsfysize=7cm
\mbox{\epsfbox{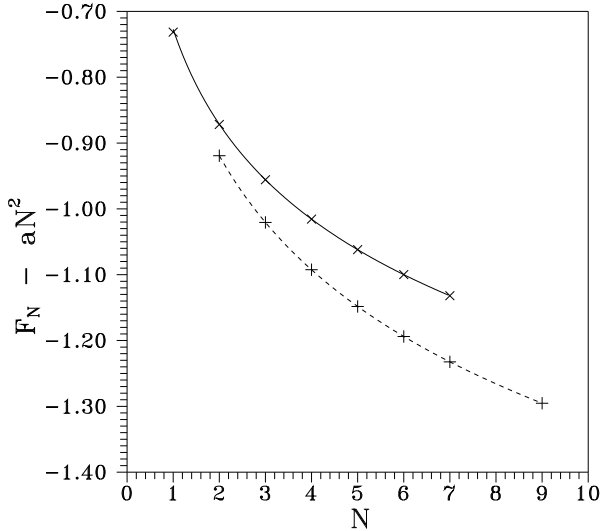}}
\end{center}  
\caption{Plot of the finite-size correction to the free energy of the
Ising model at the ferromagnetic critical point, in
lattices on the octahedron (points over the full line) and on the
tetrahedron (points over the dashed line).}
\label{f}
\end{figure}

We have computed the free energy $F_{N}$ for the members $N = 1$
up to $N = 7$ of the series of honeycomb lattices embedded on the
octahedron. The values are given in Table \ref{t1}.
We have observed a clear $\log (N)$
correction to the leading behavior $\propto N^2 $ of the free
energy as a function of the lattice size, when the
measurements are carried out at the critical coupling
$\beta_{\infty}$. 
The task is facilitated by taking into account the precise value 
of the bulk free energy per site in the honeycomb lattice,
$a/24 \approx - 0.331912$ \cite{hout}.
By computing at coupling constant $\beta =
\beta_{\infty} $ and making a sequence of fits for sets of four
consecutive lattices, we obtain the
respective values of the $b$ coefficient in (\ref{fe}),
in order of increasing lattice size : 
$-0.20486$, $-0.20763$,  $-0.20807$, $-0.20820$.
We observe a clear
convergence towards a value $b \approx -0.208 $.  We have plotted in
Fig. \ref{f} the values of $F_{N} - aN^2 $ and the best fit for
the last four points in the plot.  The sum of the squares of the
deviations from the logarithmic dependence (for $N = 2$ up to $N =
7$) is $\approx 2.7 \cdot 10^{-7} $. The accuracy of the fit is remarkable,
given that it is achieved by adjusting only 
the coefficient $b$ and the constant term in Eq. (\ref{fe}).

The above results show that the
hypothesis of finite-size scaling may be applied to the free
energy to determine the conformal anomaly on a curved
background. Let us now interpret the coefficient $b$ of the
anomaly that we have obtained for the octahedron. We assume that
the logarithmic correction can be computed as the
sum of the corrections for each of the conical singularities, $b
= \sum_{i=1}^{6} b_i$ . The coefficient $b_i$
for a conical singularity has been established in Ref.
\onlinecite{peschel} in terms
of the central charge $c$ and the angle $\theta $
enclosed by the cone:
\begin{equation}
b_i = c \frac{\theta}{24 \pi } \left( 1 - (2\pi / \theta)^2 \right)
\end{equation}
This formula leads in the case of the octahedron to
$b = -5c/12 \approx - 0.41666 \: c $,
which for $c = 1/2 $ corresponding
to the Ising model is in very good agreement with our
numerical result. This provides a
clear indication that the continuum limit of the Ising model on
the octahedron is given by a conformal field theory, with the 
same central charge as for the model in flat
space.

We move now to the family of honeycomb lattices embedded on the
tetrahedron. We may distinguish between the ferromagnetic and
the antiferromagnetic regime, since the lattices are frustrated
in this case. The finite-size scaling is actually different in
the two regimes. The number of lattice sites is given now by the
formula $n = 12 N^2$, where $N$ is the integer that labels the
member in the family. At the ferromagnetic critical coupling
$\beta_{\infty} > 0$, we have measured the free energy with the
same precision as before, for $N = 2$ up to $N = 9$. The values
that we have obtained are given in Table \ref{t2}. The
accuracy of the fit to a $\log (N)$ correction added to
the leading behavior is again remarkable. By making a
sequence of fits, each of them for four consecutive lattices, we
obtain the respective estimates for the $b$ coefficient, 
in order of increasing lattice size: $-0.2499801$, $-0.2499945$, 
$-0.2499978$, $-0.2499992$. The plot
of the function $F_{N} - a N^2 $ is given in Fig. \ref{f},
together with the logarithmic dependence from the last fit.

The result that we obtain for the coefficient $b$ is again in
very good agreement with the value expected for a conformal
field theory. The outcome of adding the effect of four conical
singularities with enclosed angle $\theta = \pi $ yields the
prediction $b = -c/2 $. Therefore, we may conclude that the
critical point in the ferromagnetic regime provides an example
of a $c = 1/2$ conformal field theory on the curved background.

\begin{figure}
\begin{center}
\epsfysize=7cm
\mbox{\epsfbox{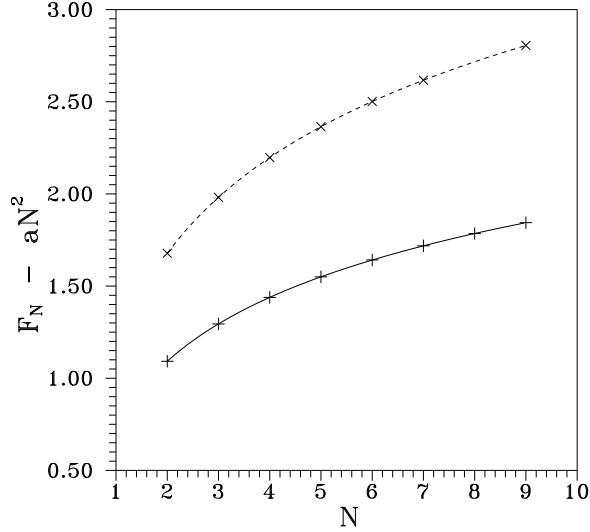}}
\end{center}
\caption{Plot of the finite-size correction to the free energy of
a free scalar field (points over the full line) and of the Ising model
at the antiferromagnetic critical point (points over the dashed line), in
lattices embedded on the tetrahedron.}
\label{af}
\end{figure}

The finite-size scaling works differently in the
antiferromagnetic regime. The values of the free energy
computed at $\beta = - \beta_{\infty } $ are given in
Table \ref{t2}.
The accuracy of the fits to determine the $\log (N) $
correction is as good as in the former instances, but now
the $b$ coefficient turns out to be positive. By carrying out  
the same sequence of fits as in the ferromagnetic regime, we find
the convergent series for the estimates of $b$ : 0.74944, 0.74986,
0.74995, 0.74996.
We have plotted in Fig. \ref{af} the values of $F_{N} - a N^2 $ and
the logarithmic fit for the last four points. The sum of the
squares of the deviations from the $\log (N) $ dependence ($N = 2$
to $N = 9$) is $ \approx 2.2 \cdot 10^{-7} $.

We can learn the correct interpretation of these results from a
similar feature of the finite-size data of the free scalar field
on the curved lattice. This can be described by a simple
tight-binding model, whose spectrum reproduces that of the
laplacian on the lattice\cite{tetra}. 
The partition function is computed
through the determinant of the coordination matrix, but the zero
mode has to be removed in order to obtain a nonsingular result.
As a consequence of that, the coefficient of the $\log (N) $
correction (fitted to data from $N = 2$ to $N = 9$ as shown 
in Fig. \ref{af}) turns out to be $\approx
 0.49999 $ . The
correct result in front of the logarithmic correction is obtained
by adding the regularized contribution of the zero mode, which
scales like $(1/2) \log(L^{-2}) $ after introducing the length
dimensions of the laplacian in the lattice.

The same effect
operates in the antiferromagnetic regime of the lattices on the
tetrahedron. These cannot be decomposed in two disjoint sublattices,
so that there is an intrinsic frustration which rules out perfect
antiferromagnetic order, irrespective of lattice size. The zero
mode is missing in the spectrum, and the correct conformal anomaly
of the $c = 1/2 $ conformal field theory is restablished adding
^^ ^^ by hand'' to the free energy the regularized zero mode
contribution $ \log (L^{-1}) $. 

To summarize, we have checked that the continuum limit of the
Ising model taken along lattices embedded on the tetrahedron and
the octahedron corresponds to respective $c = 1/2$ conformal
field theories.
We have seen that the convergence to the critical
coupling is sensibly accelerated with respect to a flat geometry
when performing the finite-size scaling in the curved lattices.
Our construction may be useful to determine the central charge 
corresponding to other models whose underlying
conformal field theory is not known.

\begin{table}
\centering
\begin{tabular}{|c|c|c|}
  $N$    & $\beta_{N}$   &   $F_{N}(\beta_{\infty}) $  \\  \hline
  1    &    0.557109(1)     &   $-8.69737867937620(1)$     \\
  2    &     0.629927(1)  &   $-32.7353289357422(1)$    \\
  3    &    0.644793(1)  &    $-72.6487548788477(1)$  \\
 4     &  0.650325(1)     &  $-128.469770730314(1)$     \\   
5      &  0.653016(1)     &   $-200.209189542789(1)$   \\
6      &  0.654540(1)    &  $-287.871899726545(1)$    \\
7      &  0.655491(1)    &  $-391.460522363036(1)$    \\
\end{tabular}

\caption{Respective values of the pseudocritical couplings $\beta_{N}$
and the free energy $F_{N}(\beta_{\infty})$ for the
octahedron.}
\label{t1}
\end{table}

\begin{table}
\centering
\begin{tabular}{|c|c|c|}
  $N$    & $F_{N}(\beta_{\infty})$   & $F_{N}(-\beta_{\infty}) $  \\  \hline
  2    &    $-16.8509982571185(1)$     &    $-14.2539387796425(1)$     \\
  3    &     $-36.8670648949387(1)$  &   $-33.8649750683343(1)$    \\
  4    &    $-64.8195835913494(1)$  &    $-61.5298831135922(1)$  \\
 5     &  $-100.721855606886(1)$     &  $-97.2090306933016(1)$     \\
6      &  $-144.579808605568(1)$     &   $-140.884668912476(1)$   \\
7      &  $-196.396605031357(1)$    &  $-192.547317539973(1)$    \\
9      &  $-323.913609253478(1)$    &  $-319.813027662780(1)$    \\
\end{tabular}

\caption{Respective  values of the free energy $F_{N}$ at the
critical couplings of the ferromagnetic and the antiferromagnetic
regime of the tetrahedron.}
\label{t2}
\end{table}

\end{document}